\documentclass{emulateapj}

\usepackage{amsmath}
\usepackage{graphicx}
\usepackage{float}
\usepackage{amssymb}
\usepackage{color}
\usepackage{subfigure}
\usepackage{sidecap}
\usepackage{color}

\shorttitle{NUCLEOSYNTHESIS IN SUPERNOVA EXPLOSIONS TRIGGERED BY A QUARK-HADRON PHASE TRANSITION}
\shortauthors{Nishimura et al.}

\begin{document}

\title{Nucleosynthesis in core-collapse supernova explosions\\
triggered by a quark-hadron phase transition}

\author{
Nobuya Nishimura\altaffilmark{1,2},
Tobias Fischer\altaffilmark{2,3},
Friedrich-Karl Thielemann\altaffilmark{1,2},\\
Carla Fr\"{o}hlich\altaffilmark{4},
Matthias Hempel\altaffilmark{1},
Roger K\"{a}ppeli\altaffilmark{1},
Gabriel Mart\'{i}nez-Pinedo\altaffilmark{2,3},\\
Thomas Rauscher\altaffilmark{1},
Irina Sagert\altaffilmark{5},
and
Christian Winteler\altaffilmark{1}}

\affil{
e-mail: nobuya.nishimura@unibas.ch\\
\altaffilmark{1}
Department of Physics, University of Basel, CH-4056 Basel, Switzerland\\
\altaffilmark{2}
GSI, Helmholtzzentrum f\"ur Schwerionenforschung GmbH, D-64291 Darmstadt, Germany\\
\altaffilmark{3}
Technische Universit{\"a}t Darmstadt, D-64289 Darmstadt, Germany\\
\altaffilmark{4}
Department of Physics, North Carolina State University, NC 27695, USA\\
\altaffilmark{5}
Department of Physics and Astronomy, Michigan State University, MI 48824, USA
}

\begin{abstract}
We explore heavy-element nucleosynthesis in the explosion of massive stars
which are triggered by a quark--hadron phase transition
during the early post-bounce phase of core-collapse supernovae.
The present study is based on general-relativistic radiation hydrodynamics simulations
with three-flavor Boltzmann neutrino transport in spherical symmetry,
which utilize a quark-hadron hybrid equation of state
based on the MIT bag model for strange quark matter.
The quark--hadron phase transition inside the stellar core forms
a shock wave propagating toward the surface of the proto-neutron star.
This shock wave results in an explosion and ejects neutron-rich matter
from the outer accreted layers of the proto-neutron star.
Later, during the cooling phase, the proto-neutron star develops a proton-rich neutrino-driven wind.
We present a detailed analysis of the nucleosynthesis outcome
in both neutron-rich and proton-rich ejecta and compare our
integrated nucleosynthesis with observations
of the solar system and metal-poor stars.
For our standard scenario we find that a ``weak" $r$-process occurs
and elements up to the second peak ($A \sim 130$)
are successfully synthesized.
Furthermore, uncertainties in the explosion dynamics could barely allow
to obtain the strong $r$-process
which produces heavier isotopes including the third peak ($A \sim 195$) and actinide elements.
\end{abstract}

\keywords{
dense matter --
nuclear reactions, nucleosynthesis, abundances --
stars: neutron --
supernovae: general
}

\maketitle

\section{Introduction}
Nucleosynthesis from core-collapse supernovae is certainly responsible
for the production of intermediate mass elements including the so-called alpha elements
and a certain fraction of iron isotopes and their neighbors (the Fe-group nuclei).
The lighter alpha elements from oxygen to silicon have contributions
from hydrostatic burning which takes place during stellar evolutions,
while the heavier ones and the Fe-group isotopes originate from explosive burning
\citep{WoosleyWeaver:1995,Thielemann:etal:1996,Woosley:etal:2002,
Nomoto:etal:2006,WoosleyHeger:2007,HegerWoosley:2010,Thielemann:etal:2011}.
A major open question is related to the source of heavy elements beyond iron.

There have been strong expectations that the innermost layer of ejecta,
close to the forming neutron star,
remains neutron-rich and is a possible site for the $r$-process.
For many years the late neutrino-driven wind (NDW), following the actual explosion,
seemed an adequate $r$-process site in core-collapse supernovae that formed a neutron star
\citep{QianWoosley:1996,Takahashi:etal:1994,Woosley:etal:1994}.
Such investigations were followed up by many parametrized calculations
that explored the sensitivity of the most relevant $r$-process
parameter, the neutron-to-seed ratio, via variations of entropy $S$,
electron fraction (or number of proton per nucleon) $Y_e$ and expansion timescale $\tau$
\citep{Hoffman:etal:1997,MeyerBrown:1997,Freiburghaus:etal:1999,Farouqi:etal:2010}.
However, steady state wind models \citep{Thompson:etal:2001,Wanajo:2006}
showed that it is very hard to attain the required entropies. These
results have been recently confirmed by fully hydrodynamical
simulations \citep{Arcones:etal:2007} that in addition showed that the
presence of a reverse shock does not have any major impact on the
neutron-to-seed ratio.

Furthermore, recent investigations noticed that the early NDW
turns matter proton-rich, producing specific Fe-group isotopes and in the subsequent
$\nu p$-process nuclei with a mass number up to $A\sim 80$ -- $90$
\citep{Liebendoerfer:etal:2003,Pruet:etal:2005,Froehlich:etal:2006a,
Froehlich:etal:2006b,Pruet:etal:2006,Wanajo:2006}. 
While it was initially hoped that there
exists a chance that the late NDW still turns neutron-rich
(after its initial, early proton-rich phase) existing core-collapse calculations
seem to indicate that the wind becomes even more proton-rich in the long-term evolution
\citep{Fischer:etal:2010,Huedepohl:etal:2010,Fischer:2012},
although very recent investigations \citep{MartinezPinedo:2012,Roberts:2012}
related to the spectral evolution of $\nu_e$ and $\bar\nu_e$
in medium might change this somewhat.

It was realized that the best chance to eject neutron-rich
matter, is to utilize material stemming from the initial collapse and compression, where
electron captures turned it neutron-rich early in the explosion, before
neutrino interactions have the chance to convert it into proton-rich matter in the NDW.
Electron-capture supernovae, which explode without a long phase of
accretion onto the proto-neutron star, apparently provide such conditions
\citep{Wanajo:etal:2011}.
However, the $Y_{e}$'s obtained under such conditions do not
support a strong $r$-process, which successfully reproduces the platinum
peak of $r$-elements around $A=195$.

An alternative explosive scenario as a possible site for the
$r$-process has been suggested by \citet{Jaikumar:2007}, where after 
a supernova explosion a quark--hadron phase transition inside the neutron
star can eject very neutron-rich material from the neutron star surface.
This scenario has been referred to as a \emph{quark-nova}.
In the present paper, we follow a different approach and
investigate core-collapse supernova explosions that are triggered by
a quark-hadron phase transition during the early post-bounce
phase \citep[for details, see][]{Sagert:etal:2009,Fischer:etal:2011}
and their nucleosynthesis features.
Under such conditions, zones which are initially neutron-rich, can be
(promptly) ejected without experiencing the strong effects of the neutrino
flux which comes from the central proto-neutron star.

In the following sections, we discuss the supernova explosion model 
and the nuclear physics inputs utilized for the nucleosynthesis calculations
in Section 2,
the conditions experienced and the resulting ejecta in Section 3,
and give a detailed analysis of the ejecta composition in Section 4,
followed by conclusions, also in comparison to alternative $r$-process sites,
in Section 5.
%

\section{Methodology and nuclear physics input}
We explore nucleosynthesis in core-collapse supernovae
which are triggered by a deconfinement phase transition during the early post-bounce phase.
The explosion models have been discussed in detail by \citet{Fischer:etal:2011}.
In this section, we summarize the main features of the explosion models
and the input physics of the nuclear reaction network utilized for nucleosynthesis.

\subsection{Hydrodynamics and Equation of State}
A self-consistent supernova explosion model, which we adopt to investigate nucleosynthesis
in the present work, has been carried out by using the AGILE-BOLTZTRAN code.
This numerical code is based on general-relativistic radiation hydrodynamics in spherical symmetry
with an adaptive grid and employs a three-flavor Boltzmann neutrino transport
\citep{Liebendoerfer:etal:2004} with detailed microphysics,
e.g., a realistic nuclear equation of state (EOS),
weak processes, and (other) nuclear reactions.
For the current study, the standard weak processes were considered
as listed in Table 1 of \citet{Fischer:etal:2011}.

The baryonic EOS in supernovae needs in general to be known
for the following three intrinsically different regimes:
\begin{enumerate}
\item At temperatures below $6$ GK ($\simeq 0.5$ MeV),
the baryon EOS is dominated by heavy nuclei and their abundances,
determined by individual nuclear reactions which are not necessarily in equilibrium.
In \citet{Fischer:etal:2010}, we included a nuclear reaction network, consisting of 20 nuclei,
which permits to simulate a large domain of the progenitor up to the helium--layer
and is utilized to predict the amount of energy generation by explosive nuclear burning.
\item At temperatures above $6$ GK ($\simeq 0.5$ MeV),
the nuclear statistical equilibrium (NSE) can be applied,
where the baryonic EOS from \citet{LattimerSwesty:1991} and \citet{Shen:etal:1998}
are commonly used in supernova simulation studies.
\item Above the nuclear matter density, the state of matter is highly uncertain.
There exists the possibility of a deconfinement phase transition.
Therefore, we extended the hadronic EOS by \citet{Shen:etal:1998} at high densities and temperatures,
making use of a quark EOS based on the bag model for strange quark matter.
For the first-order phase transition between hadronic and quark phases,
we applied Gibbs conditions leading to a mixed phase during the transition.
This results in a continuous phase transformation.
Details of the quark--hadron hybrid EOS are discussed in \citet{Fischer:etal:2011}.
\end{enumerate}
Besides the contributions from electrons and positrons as well as photons, Coulomb corrections
to the EOS are added by the method of \citet{TimmesArnett:1999}.

For the current nucleosynthesis predictions, we select the explosion calculations
of the $10.8$ $M_\odot$ progenitor model, where a quark--hadron hybrid EOS
was used with an early phase transition to quark matter close to normal nuclear
matter density \citep[labeled EOS2, second line of Table 2 in][]{Fischer:etal:2011}.
We chose this model because it has an explosion energy consistent
with the expected order of magnitude of $10^{51}$ erg
\citep[see the second line of Table 3 in][]{Fischer:etal:2011}.
The maximum gravitational mass for this EOS is $1.5026$ $M_\odot$.
Though it is in agreement with the highest precisely known mass
of a compact star, the Hulse--Taylor pulsar of $1.44$ $M_\odot$,
recent mass limits for the physical EOS are based on the millisecond pulsars J1903+0327
with $M = 1.667 \pm 0.021$ $M_\odot$
\citep{Freire:etal:2011} and J1614--2230 with a high mass of
$M = 1.97 \pm 0.04$ $M_\odot$ \citep[][]{Demorest:etal:2010}.
The inclusion of corrections from the strong interaction coupling constant
can stiffen the quark EOS and lead to higher maximum masses
\citep[see, e.g.,][]{Schertler00,Alford:etal:2005,Sagert:etal:2010,Sagert:etal:2011,Weissenborn:etal:2011}. 

The physical conditions for a possible quark-hadron phase
transition in the proto-neutron star are highly uncertain.
Therefore, \citet{Fischer:etal:2011} also constructed a quark-hadron hybrid
EOS which includes corrections from the strong interaction coupling constant.
They showed that such an EOS, with a maximum mass of 1.67~M$_\odot$
and a phase transition to quark matter close to nuclear saturation density
\citep[see EOS3 in table~2 and 3 in][]{Fischer:etal:2011},
leads to a qualitatively similar explosion scenario in spherical symmetry as
obtained in the explored models with several EOSs which have lower maximum masses.
Note, the different post-bounce times for the onset of deconfinement, and
hence for the onset of the explosion, lead to different proto-neutron star structures.
In general, a delayed phase transition at higher densities translates to a longer
mass accretion phase (for the same progenitor).
It results in a more massive protoneutron star with a steeper density gradient
at its surface and a lower $Y_e$ at the onset of the explosion triggered by the
phase transition.
An earlier phase transition and less steep density
gradient would result in a slower expansion and a slightly higher $Y_e$.

This has important consequences for the nucleosynthesis of the ejected material
since expansion timescale and enclosed mass, and also neutrino luminosities and spectra,
depend on these aspects
\citep[for a detailed discussion, see][]{Fischer:etal:2011} and the behavior
of EOS1 and EOS2 in their Figure 17.

\citet{Weissenborn:etal:2011} recently showed that it is possible to
obtain a quark--hadron hybrid EOS which allows for both,
a maximum mass larger than 2 $M_\odot$ and a low critical density for the appearance of quark matter.
Whether such quark--hadron hybrid EOS will result in a similar dynamical evolution
as discussed in \citet{Fischer:etal:2011} will be examined in future simulations.

\subsection{Explosion Scenario}
The supernova post-bounce evolution is characterized by mass accretion
causing a continuous rise of the central density.
Once the central density exceeds the critical density for the onset of deconfinement,
the quark--hadron phase transition takes place, leading to the appearance of a quark--hadron mixed phase.
Thereby, quark matter appears in the supernova core where the highest densities are experienced.
The timescale for the appearance of quark matter is given by the timescale for the central density to rise.
This depends on the progenitor model, which determines mass accretion rates, and the hadronic EOS.
Note that the EOS in the mixed phase is significantly softer than in hadronic and pure quark phases.
It is a consequence of the assumed first-order phase transition.
Mass accretion from the outer layers of the progenitor onto the central supernova core
leads to a continuous rise of the central enclosed mass.
When the critical mass (given by the hybrid EOS) of the configuration is obtained,
it becomes gravitationally unstable and the supernova core begins to contract.
The contraction proceeds into a collapse which rises the density
and converts the hadronic core into quark matter at around the center.
A massive pure quark core forms at the center, where the EOS stiffens and the collapse halts,
and an accretion shock forms.
The shock wave propagates out of the high-density supernova core,
remaining an accretion front with no matter outflow.
Once it reaches the outer layers of the central core, where the density drops over several orders of magnitude,
the accretion front accelerates and turns into a dynamic shock with matter outflow.
This moment determines the onset of an explosion,
also for supernova models which would otherwise not explode in spherical symmetry,
based on explosion mechanisms discussed so far.
Finally, at distances on the order of $100$ km the expanding shock wave merges
with the standing shock from the initial core bounce at nuclear densities,
which was unaffected by the dynamics occurring in the supernova core.

\begin{table}[htp]
\centering
\caption{Summary of mass zones and their properties}
\begin{center}
\begin{tabular}{c c c c c}
\hline
\hline
	Zone No.
	& $M_{\#}$ ($10^{-2} M_{\odot}$)
	& $\Delta\overline{M_\#}$ ($M_{\odot}$)
	& $Y_{e, \rm{NSE}}$
	& $t_{\rm{ej}}$ (s) \\
	\hline
	001--014                    &                   $0.000$--$0.208$
	& {$1.496\times10^{-4}$} & $                         0.20$
	& $\cdots$        \\
	\textcolor{red}  {015--019}  & \textcolor{red}  {$0.210$--$0.216$}
	&\textcolor{red}{$1.474\times10^{-5}$} & \textcolor{red}  {$\sim 0.55$}
	&\textcolor{red}  {$1.5$--$4.0$} \\
	\textcolor{green}{020--050} & \textcolor{green}{$0.217$--$0.232$}
	&\textcolor{green}{$1.063\times10^{-5}$} & \textcolor{green}{$\sim 0.33$}
	&\textcolor{green}{$\sim 0.5$} \\
	\textcolor{blue} {051--120} & \textcolor{blue} {$0.250$--$1.482$}
	&\textcolor{blue}{$1.786\times10^{-4}$} & \textcolor{blue} {$0.33 \sim 0.50$}
	&\textcolor{blue} {$\sim 0.5$} \\
	\hline
\end{tabular}
\end{center}
\footnotetext{$M_{\#}$: mass coordinates relative to the innermost zone of $1.48 M_{\odot}$.
$\Delta\overline{M_\#}$: the averaged mass of the zone.
$Y_{e,\rm{NSE}}$: $Y_e$ at the end of NSE (below $T = 9$ GK).
$t_{\rm{ej}}$: ejection time after the bounce.}
\label{tab-properties}
\end{table}

When the (second) shock reaches the neutrino spheres,
an additional millisecond neutrino burst is released.
It appears in all flavors, however dominated by $\bar\nu_e$ and $\nu_{\mu/\tau}$,
in contrast to the $\nu_e$-deleptonization burst
related to the early bounce shock propagation across the neutrino spheres
between $200$ and $500$ ms after core bounce.
This second neutrino burst is of particular interest for water-Cherenkov neutrino detectors,
which are more sensitive to $\bar\nu_e$ than to $\nu_e$.
\citet{Dasgupta:etal:2010} demonstrated that the currently operating generation
of water-Cherenkov neutrino detectors (e.g., Super-Kamiokande and IceCube detectors)
can resolve such millisecond neutrino burst of the explosion models by \citet{Fischer:etal:2011}.

The matter considered for nucleosynthesis studies of heavy elements (see Section 3) belongs 
originally to the inner parts of the silicon and sulfur layers of the $10.8$ $M_\odot$
progenitor from  \citet{Woosley:etal:2002}, with temperature and electron fraction of
$3$ GK and $Y_e \simeq 0.5$, respectively, at $800$ -- $1000$ km from the center
on the pre-collapse phase.
During the collapse and the explosion, material contracts and is heated by the shock
to temperatures exceeding $100$ GK and hence completely dissociates into free nucleons.
At high densities weak processes, mainly electron captures, establish a very low proton-to-baryon
ratio $Y_e \simeq 0.1$ during the core-collapse.
The further evolution is discussed in detail in Section 3.

\subsection{Nuclear Reaction Network}
The nuclear reaction network utilized
for the following nucleosynthesis simulations of the ejecta
is an extension of previous ones, which
have already been described in detail \citep[see][]{Nishimura:etal:2006,Fujimoto:etal:2008}.
The network includes more than $4000$ nuclei from neutrons and protons up to 
fermium with atomic number $Z = 100$
\citep[for detail, see Table 1 in][]{Nishimura:etal:2006}
and includes proton-rich isotopes as well as neutron-rich ones far from stability.
It includes two- and three-body reactions, decay channels, and
electron as well as positron capture 
\citep[for details, see network A in][]{Fujimoto:etal:2007}
and screening effects for all relevant charged-particle reactions.
Experimentally determined masses \citep{AudiWapstra:1995}
and reaction rates are adopted if available.
Otherwise, theoretical predictions for nuclear
masses, reaction rates, and beta-decays are applied, based on
the finite range droplet mass model
\citep{Moller:etal:1995}.
Spontaneous and beta-delayed fission processes \citep{StaudtKlapdor-Kleingrothaus:1992}
are taken into account.
We adopt the empirical formula for fission fragments by \citet{KodamaTakahashi:1975}.

\begin{figure}[htbp]
	\begin{flushleft}
		\includegraphics[width=\hsize]{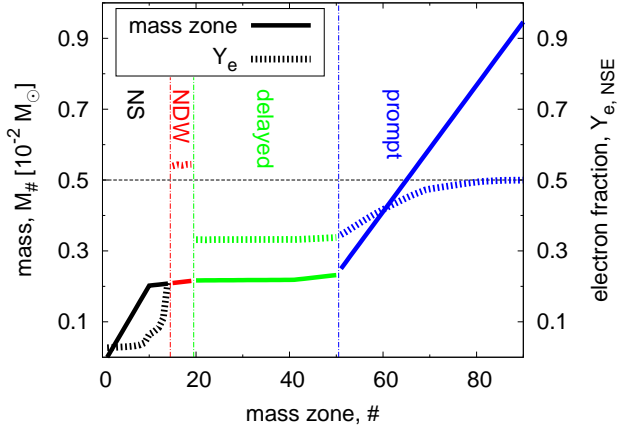}
		\caption{
		Initial distribution of mass and $Y_e$ as a function of mass zone number.
		Thick lines and dashed lines relate to Lagrangian mass coordinate
		and electron fraction, respectively.
		Masses are measured from the surface of the proto-neutron star
		(starting at 0 for zone 001 of Table \ref{tab-properties})
		and the electron fractions are adopted at the time
		when the temperature decreases down to $T=9\times 10^9$~K for ejected matter
		and the end of the hydrodynamic simulation for inner non-ejected zones).
		The plot shows only the mass zone range 001 to 090 for enclosed mass and $Y_e$.
		For the mass zones 90 to 120, $Y_e$ stays essentially constant at $\simeq 0.5$ and the
        enclosed mass continues to be proportional to mass zone numbers.}
	\label{fig-init}
	\end{flushleft}
\end{figure}

We also employ neutrino interactions with matter in order to include
dominant weak interactions affecting the evolution of the overall
proton/nucleon ratio $Y_{e}$.
For (anti-)electron neutrino captures by nucleons,
we adopt reaction rates derived by \citet{QianWoosley:1996}
but ignore any reactions with heavy isotopes
because the amount of neutrino capture is negligible for the early phase of nucleosynthesis.
The neutrino fluxes, resulting from the detailed neutrino-radiation hydrodynamics
calculation described in the previous subsection,
are utilized to determine the actual rates as a function of time.
These reaction rates depend on the distance from the proto-neutron star and the
mean energy and luminosity of neutrinos emitted from the proto-neutron star.
It depends on the structure and the evolution, which is sensitive to the EOS,
and the precise evolution history of ejected matter including the early phase of the core bounce.
Thus, this is different from the treatment of other nuclear reaction rates,
which are determined only by local thermodynamic conditions and density and 
temperature.

\section{Nucleosynthesis in the ejecta}

\subsection{Dynamic Evolution of Mass Zones}
In order to calculate the nucleosynthesis evolution of ejected matter
within a postprocessing approach, the dynamic evolution is required in
radial Lagrangian mass zones.
For this reason, the evolution determined with
the radiation hydrodynamics code AGILE-BOLTZTRAN 
\citep[see][]{MezzacappaBruenn:1993a, MezzacappaBruenn:1993b,
MezzacappaBruenn:1993c,Liebendoerfer:etal:2001a,Liebendoerfer:etal:2001b}
which is based on an adaptive grid,
was mapped on a Lagrangian grid of 120 mass zones.
This provides the Lagrangian evolution of physical quantities,
such as density, temperature, electron fraction, and velocity of the ejected material,
and in addition the neutrino fluxes experienced as a function of time.

The mass zones which are ejected in the explosion are classified in three different categories,
related to their ejection process and thermodynamic quantities.
As listed in Table {\ref{tab-properties}}, zones 001 to 120,
given with the ejection timescale after bounce ($t_{\rm{ej}}$) and the final
$Y_{e,\rm{NSE}}$ which is $Y_e$ at the end of NSE (below $T=9$ GK),
cover the material from the surface of the inner core at 1.48000 $M_{\odot}$
to layers with a corresponding mass of 1.49482 $M_{\odot}$. They are also shown with respect 
to their mass as well as $Y_e$-distribution in Figure \ref{fig-init}.

\begin{figure}[htbp]
	\begin{center}
		\includegraphics[width=\hsize]{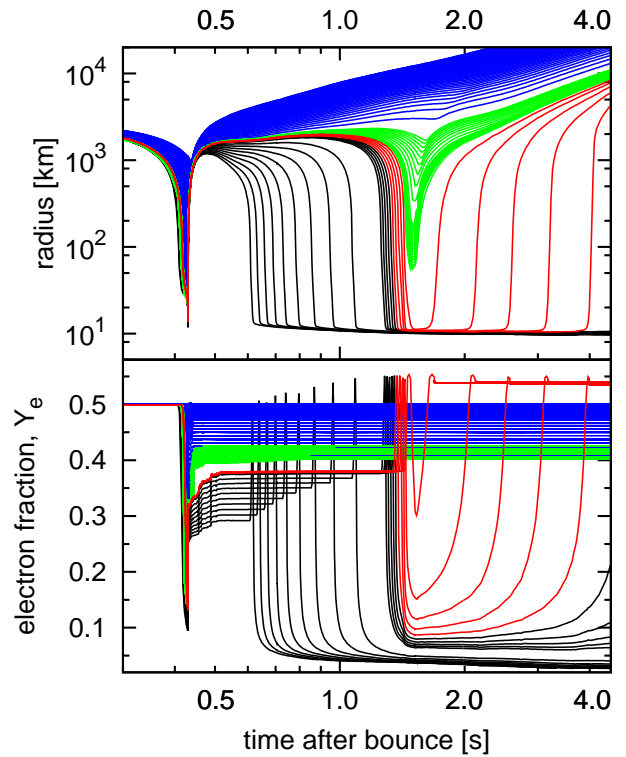}
		\caption
	{Top:
	radial trajectories of mass elements as a function of time after bounce.
	The colors indicate the properties of these mass elements:
	black, red, green and blue lines refer to matter which is either (black) not ejected,
	(red) part of the neutrino-driven wind, (green) initially stalled 
	matter which gets boosted by the wind and (blue) matter which experiences a 
	prompt ejection.
	Bottom: evolution of $Y_e$ as a function of time after the core bounce.
	The deconfinement phase transition happens about
	0.4 s after this initial bounce, causing the explosion and ejection of matter.
	The colors are the same as in the top panel.}
	\label{fig-properties}
	\end{center}
\end{figure}

These zones coincide with the matter discussed at the end of Section 2.2,
where at high densities a $Y_e$ decrease down to even $0.02$ has been noticed.
The evolution of these mass zones in time is displayed in 
Figure {\ref{fig-properties}}.
Zones 001 to 014 are not ejected within $0.5$ s after the core bounce.
These zones preserve the original low-$Y_e$ obtained during collapse
and shock wave propagation.
As they are not ejected, we ignore them in the further nucleosynthesis discussion,
plus all matter originating from regions at smaller radii.
In Figure {\ref{fig-properties}}, they are displayed in black.
Zones 015 to 019 are ejected in the so-called NDW, shown in red.
Their $Y_e$ is strongly affected by neutrino interactions, turning this matter proton-rich.
Zones 020 to 050 (displayed in green) have stalled from infall
after shock formation and are ejected thereafter
due to neutrino heating and dynamic effects \citep{Fischer:etal:2011}. 
The adjacent zones 051 to 120 are ejected in a prompt way, due to the shock wave
originating from the deconfinement phase transition (displayed in blue).
We clearly see the division of matter which is ejected in a prompt fashion (blue), 
matter which is coasting and falling in again, but 
gets reaccelerated outward by neutrino energy deposition (green), matter which
falls back onto the neutron star, but becomes part of the NDW ejecta
(red), and finally matter which stays on the neutron star and will never
be ejected (black). 

\begin{figure}[htbp]
	\begin{center}
		\includegraphics[width=\hsize]{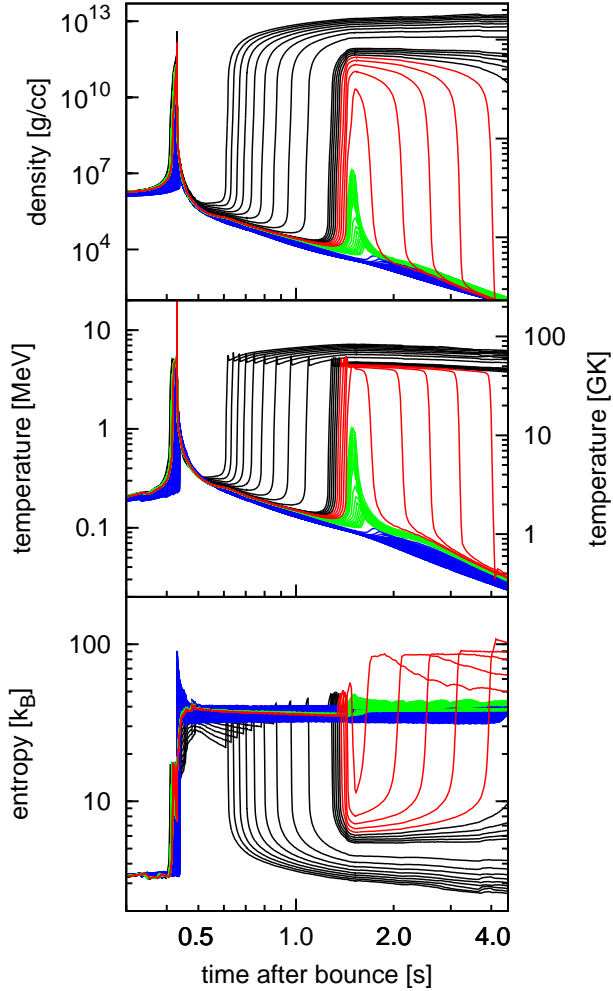}
		\caption{Evolution of temperature (top), density (middle),
		and entropy (bottom) of mass zones as a function of time after bounce. }
\label{fig-postprocess}
\end{center}
\end{figure}

\begin{figure}[htbp]
	\begin{center}
		\includegraphics[width=\hsize]{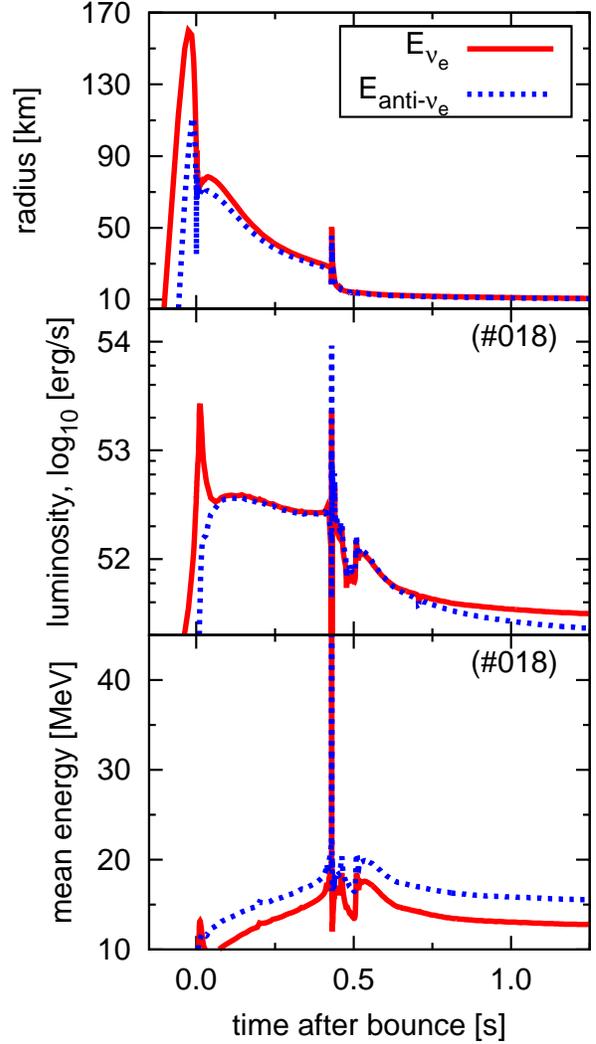}
\caption{Radius of neutrino spheres for electron neutrinos and electron anti-neutrinos (top).
Luminosities (middle) and mean energies (bottom) experienced
by mass zone 018, which belongs to the neutrino-driven wind with proton-rich ejecta,
as a function of time after bounce. The difference between the mean energy of anti-neutrinos and 
those of neutrinos is of the order $3$ MeV, i.e., less than $4\Delta$, where $\Delta$ is the
neutron--proton mass difference.}
\label{fig-neutrinos}
	\end{center}
\end{figure}

As shown in the bottom part of Figure {\ref{fig-properties}},
the blue and green zones are neutronized during the collapse
via electron capture to various degrees,
depending on the maximum density attained.
Thus the main feature is that a strong compression during infall leads to high densities
(at still low entropies, i.e. highly degenerate matter)
with large electron Fermi energies which endorse electron captures and a strong
neutronization of matter with a small $Y_e$.
Their $Y_e$-values range from $0.35$ to $0.5$.
The prompt or quasi-prompt ejection does not change this value
(with minor effects on the innermost zones, being partially affected by the NDW).
The effect of neutrino and anti-neutrino exposure from the core leads
also to an increase of $Y_e$, as discussed before.
Therefore, the major point in favor of an $r$-process
is a fast expansion of that material having two aspects:
(1) timescales are shorter than needed for attaining weak equilibrium and
(2) matter moves fast to larger radii where the effect of neutrinos vanishes ($1/r^2$).
The mass zones in red experience a similar effect in their early evolution
during collapse, but the later evolution leads to values of $Y_e$ 
exceeding $0.5$. This is similar to recent studies of the NDW 
\citep[see the Introduction and][]{Fischer:etal:2010}, where similar neutrino 
and anti-neutrino spectra and
flux intensities favor proton-rich matter due to the neutron--proton mass
difference, resulting in different energies available for the 
neutrino/anti-neutrino captures. As the neutrino luminosity is still high
in the ejection phase, we expect $\nu p$-process nucleosynthesis.
Finally non-ejected mass zones (black) can initially also experience interaction with the neutrino flux
and turn proton-rich while still at larger radii and small(er)
densities. Once they settle on the surface of the neutron star at high 
densities, capture of degenerate electrons dominates over the neutrino effects,
and they turn neutron-rich again.

\begin{figure}[htbp]
	\begin{center}
		\includegraphics[width=\hsize]{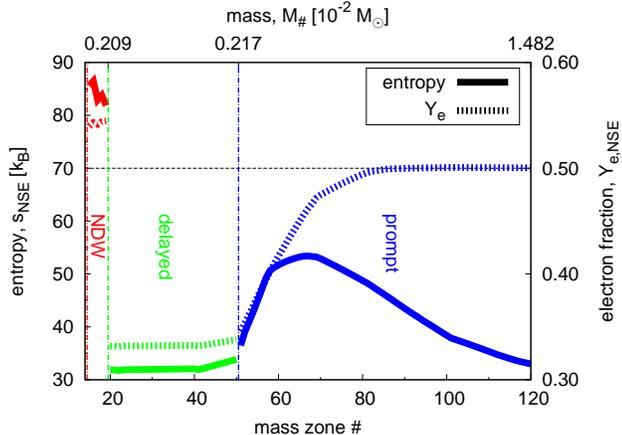}
		\caption{
$Y_e$ and entropy $S$, which set the conditions for explosive nucleosynthesis
at the time of matter ejection. The innermost ejected zones are proton-rich due
the effect of the neutrino-driven wind, which also heats matter efficiently, leading
to high entropies. The outer mass zones, ejected in a more prompt fashion
keep their original (slightly neutron-rich) $Y_e$ from the infall/compression
phase. }
\label{fig-entYe}
\end{center}
\end{figure}

The thermodynamic conditions (density, temperature, and entropy),
which are responsible for the nucleosynthesis results, are shown in 
Figure {\ref{fig-postprocess}} for the ejected mass zones. These figures indicate 
a temperature, density, and entropy maximum when the quark--hadron phase 
transition occurs, which causes a second core bounce
(about $0.4$ s after the first bounce at nuclear densities; see Figure \ref{fig-properties})
and an outgoing shock front forms.
The expansion follows a close to constant entropy, i.e. is adiabatic,
once matter is ejected (after $0.4$ s for the prompt ejection and after $1.5$ s for the
delayed ejection). The matter which initially fell back onto the proto-neutron star
and is finally ejected by the NDW, experiences heating and an 
entropy rise due to this energy deposition by neutrinos.

\subsection{Neutrinos from the Proto-neutron Star}
The properties of the neutrino and anti-neutrino flux (luminosities, average 
energies and neutrino sphere radii) can be found in Figure {\ref{fig-neutrinos}.
It is clearly seen that the initial bounce ($0$ s) at nuclear densities leads to 
a neutrino burst due to electron captures, while the second shock wave 
caused by the quark-hadron phase transition (at $0.4$ s) also produces antineutrinos.
From that point on in time the neutrino and anti-neutrino luminosities are
comparable (slightly smaller for anti-neutrinos). The average energies are
larger for anti-neutrinos than for neutrinos, but the difference remains less
than 4 MeV. Neutrino and anti-neutrino captures determine the neutron/proton
ratio due to the reactions
$$\bar\nu_e + p \rightarrow n + e^+ \ \ \ \ \ \nu_e + n \rightarrow p + e^-.$$

\begin{figure}[htbp]
	\begin{center}
		\includegraphics[width=\hsize]{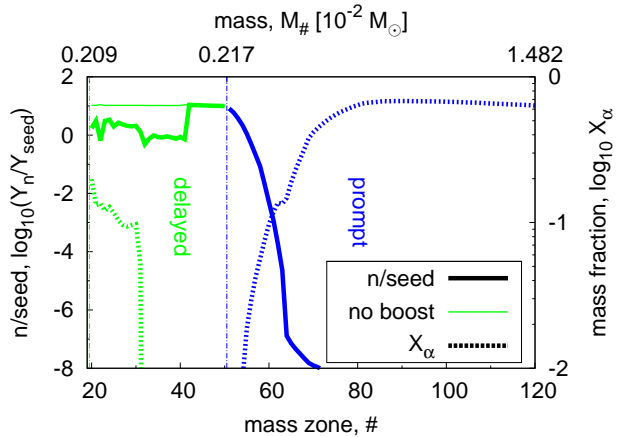}
		\caption{
	Neutron/seed ratio and remaining mass fractions of $^4$He ($\alpha$-particles)
	after charged-particle freeze-out.
	Both properties result from the original $Y_e$ and entropy $S$ in these mass zones.
	In turn they determine the fraction of heavy elements and whether those experience
	further neutron capture after charged-particle freeze-out,
	which is the key to the pattern of heavy nuclei and the maximum mass number attained.
	In the inner part (green lines) an effect is seen, which results from an intermediate fallback
	before final ejection. These mass zones experience first (due to the shock from the deconfinement
	phase transition) a maximum temperature and density, expand afterwards close to adiabatically, heat
	up during the intermediate fallback, and then expand freely. The line indicated with ''no-boost'' gives
	the initial neutron/seed ratio after the first expansion. The alpha-fraction in the inner zones
	results from the reheating phase. The initial expansion at low $Y_e$'s would result in vanishing
	alpha-fractions.}
\label{fig-Xalnseed}
\end{center}
\end{figure}

Based on the neutron/proton mass difference of $\Delta = 1.293$~MeV,
\citep{Froehlich:etal:2006a} could show (see their Equation (4)),
that with the use of Equations (64a) and (64b) in \citet{QianWoosley:1996},
$\dot Y_e > 0$ in the case
that the difference between the mean antineutrino and neutrino energies fulfills
$\epsilon_{\bar{\nu}} - \epsilon_\nu < 4\;\Delta$,
with $\epsilon = \langle E^2\rangle/\langle E \rangle$ and where
$\langle E \rangle$ is the mean energy and $\langle E^2 \rangle$ is the square
value of the root-mean-square (rms) energy.
Therefore $Y_e > 0.5$ is obtained if the timescale for
neutrino/anti-neutrino captures is shorter than the dynamic timescale.
Thus, for all conditions discussed here, where neutrino and anti-neutrino
captures are responsible for the $n$/$p$ ratio, proton-rich conditions are attained,
i.e., $Y_e > 0.5$.
That is exactly what is seen in Figures \ref{fig-properties}
and \ref{fig-neutrinos} for mass zones which experience essential
neutrino fluxes (weighted by $1/r^2$) at radii of about 100 km.
It is related to the spectral evolution of $\nu_e$ and $\bar\nu_e$ after
the onset of explosion and has been discussed in detail in \citet{Fischer:2012}.
For matter at  larger radii (about 1000 km), the timescale for this process is
too long and minor $Y_e$ changes occur, i.e., the initial $Y_e$
from the collapse phase is retained.
Matter at smaller radii, on top of the neutron star, experiences high densities
and  electron Fermi energies, where electron captures dominate which make matter
neutron-rich. 

Figure \ref{fig-entYe} underlines this effect due to neutrino
interactions or electron capture. All outer mass zones keep their original
$Y_e$, which is due to electron capture at high densities during the collapse, 
and ranges from about $0.48$ (further out) to $0.32$ (for the inner quasi-prompt
ejected matter). Material which fell in initially onto the surface of the 
neutron star and is then ejected via the NDW, has been turned proton-rich 
by neutrino and anti-neutrino captures with values up to $Y_e=0.55$.
The NDW also leads to energy 
deposition and an entropy increase to maximum values of about $85$ $k_B$ per
baryon. The entropy in the outer ejected regions is of the order $30$--$50$ $k_B$
per baryon, caused by shock heating during the passage of the ejection
shock wave.

\subsection{Nucleosynthesis Results}

In the following we show final nucleosynthesis results for a number of typical mass zones.
In order to get a rough idea about the results of explosive nucleosynthesis,
one can utilize either maximum densities and temperatures prior to an adiabatic expansion
or the entropies attained in the expanding matter (in radiation-dominated regimes $S\propto T^3/\rho$).

\begin{figure*}[htp]
\begin{center}
	\includegraphics[width=0.7\hsize]{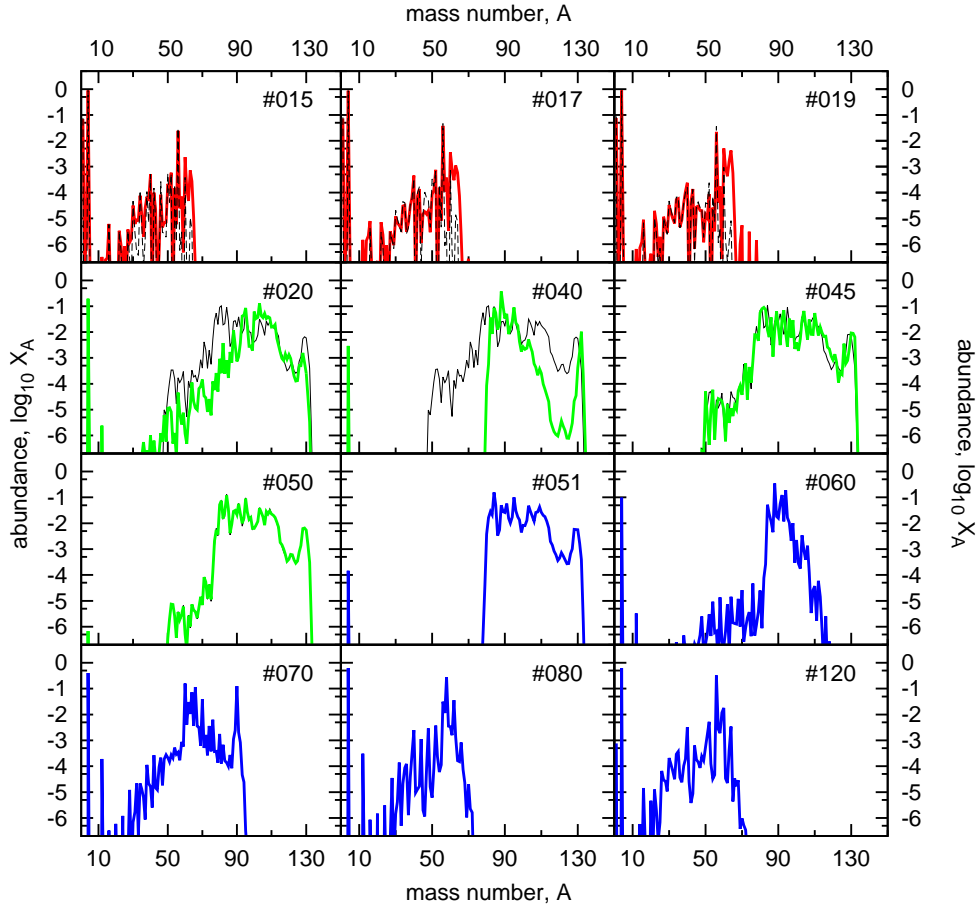}
	\caption{Final nucleosynthesis results for selected mass zones
	showing mass fractions as a function of nuclear mass number A.
	The mass zone number is given in each panel. In panels, where we 
	show dashed and solid lines, the solid lines correspond
	always to the final result. In zones up to 19 the dashed lines
	correspond to abundances without including the $\nu$p-process, i.e.
	before proceeding beyond the beta-decay bottle-neck $^{64}$Ge. In zones
	20-45 they show abundances before the reheating boost.}
		\label{fig-finab}
\end{center}
\end{figure*}

Comparing entries in Figure 5 of \citep{Thielemann:etal:1990} and Fig.3 of 
\citep{Thielemann:etal:1996} leads to the conclusion that (1) these are
typical conditions for an alpha-rich freeze-out from explosive Si-burning
and (2) one would expect remaining alpha mass-fractions after charged-particle
freeze--out of the order 20\% -- 100\%. One should consider, however, that 
those calculations were performed for hydrodynamic (i.e., free fall) expansion timescales,
which can differ from the actual simulation, and a value of $Y_e=0.4988$, i.e.
matter neither neutron nor proton-rich. Therefore we expect the following
changes: (1) higher/lower entropies within the given variety will lead to 
higher/lower remaining alpha-fractions; (2) higher $\dot{Y}_e$, i.e., more proton-rich
matter causes an alpha-rich charged-particle freeze-out with remaining free
protons, which can thereafter lead to a $\nu p$-process, if a sufficient flux
of electron anti-neutrinos is still present; (3) smaller $Y_e$'s, i.e., more 
neutron-rich matter permits to bypass the slower triple-alpha reaction
via the faster $\alpha\alpha n$-reaction, in order to produce heavier nuclei
and a reduction in the remaining alpha-fraction is expected. In addition, free
neutrons are remaining after the charged-particle freeze-out.
With respect to this latter aspect, we expect also the additional behavior:
higher entropies and lower $Y_e$'s lead to a larger neutron/seed ratio,
seed nuclei being the heaviest nuclei formed after charged-particle freeze-out,
and permit therefore more neutron captures on these seed-nuclei.
Dependent on the neutron to seed ratio, this could lead to light, medium or
strong $r$-processing, producing nuclei in the first, second or third $r$-process
peaks, around A=80, 130 or 195, depending on the actual n/seed ratio attained. 

\begin{figure}[htbp]
	\begin{center}
		\includegraphics[width=\hsize]{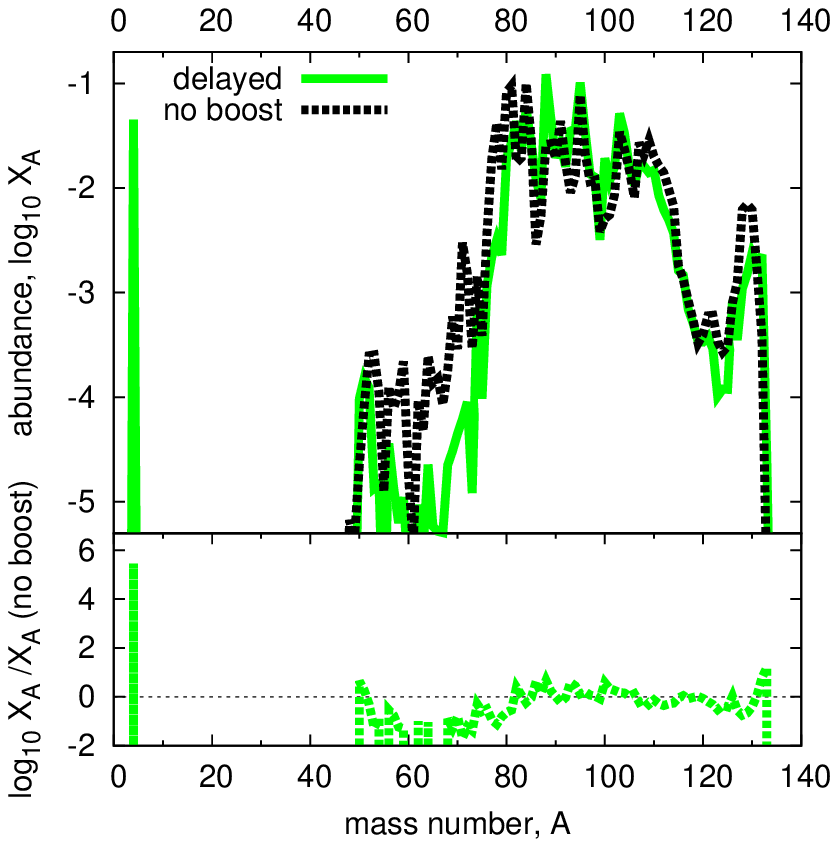}
		\caption{\textbf{Top:}
		The comparison of results with (solid) and without (dashed) the effect of reheating
		and second expansion (boost) integrated over the whole range of mass zones
		20-50.
		\textbf{Bottom:} The ratios of the two cases, based on the full hydrodynamic evolution
		with reheating (boost) and the neglection of reheating around $1.5$ s.The reheating leads to
		photodisintegrations, a strong appearance of $^{4}\rm{He}$ and a reshaping of 
		the heavy element distribution.}
\label{fig-boost}
\end{center}
\end{figure}

\begin{figure}[htbp]
	\begin{center}
		\includegraphics[width=\hsize]{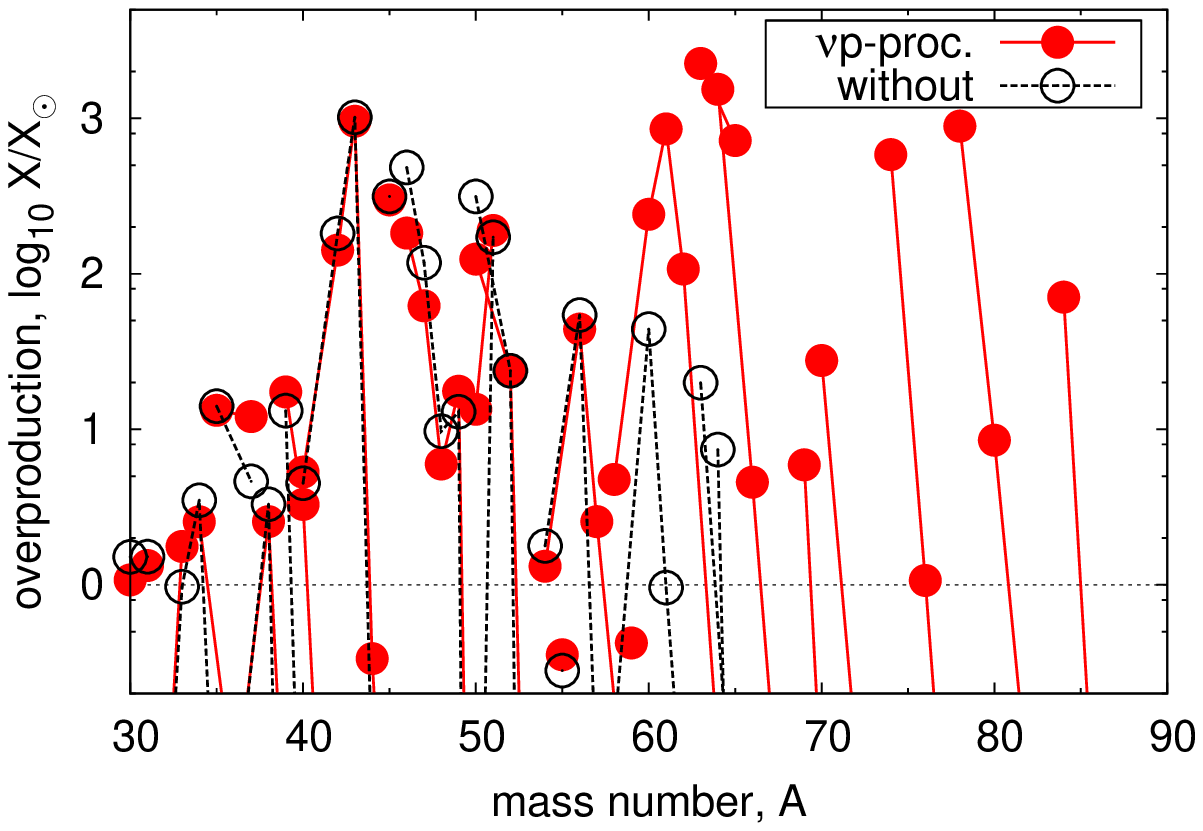}
		\caption{
Resulting nucleosynthesis for the entire mass zones in the neutrino-driven wind,
experiencing a $\nu$p-process.
The result is shown for two options:
(a) including the anti-neutrino captures which turn protons into neutrons in the late 
phase of nucleosynthesis and permit to overcome the $^{64}$Ge beta-decay bottle neck
via an (n,p)-reaction (filled red circles),
(b) neglecting this effect (open black circles).
It can be seen that the inclusion of anti-neutrino
reactions enhances abundances of nuclei heavier than $A = 64$ 
and permits the production of nuclei up to $A = 80 - 90$.}
\label{fig-nup}
\end{center}
\end{figure}

In Figure \ref{fig-Xalnseed} the properties of mass zones 020 to 120 are summarized,
those with a $Y_e < 0.5$ which experience neutron-rich conditions to a varying degree.
Properties (1), i.e., the degree of alpha-rich freeze-out and (3), the resulting 
neutron/seed ratio, are displayed. We see a complex dependence of these properties
on entropy $S$, $Y_e$, expansion timescale $\tau$, plus further complications from reheating 
and re-expansion which
do not fit to a simple expansion interpretation.
For similar $Y_e$'s, the remaining alpha-fraction
increases with entropy, as expected, when looking at the behavior of mass zones 080 to 120.
Then, following mass zones further in, the decrease in $Y_e$ dominates, 
which permits to pass the alpha-to-carbon 
bottle-neck more efficiently, via the $\alpha\alpha n$-reaction, and the remaining
alpha-fraction vanishes. As known from
moderately neutron-rich NDW simulations, none of the entropies encountered
here leads to sizable neutron/seed ratios.
For such low entropies, only a strongly neutron-rich initial composition
permits large(r) neutron/seed ratios.
This is what can be noticed when following mass zones from 070 down to 040,
where the lowest $Y_e$'s are encountered.
Mass zones 020 to 040 experience a more complicated history,
initial expansion after shock passage, later partial fallback, and then final ejection
and further expansion. Here, the neutron/seed ratio after the first charged-particle
freezeout is the one of importance for the production of heavy elements.

\begin{figure}[htbp]
	\begin{center}
		\includegraphics[width=\hsize]{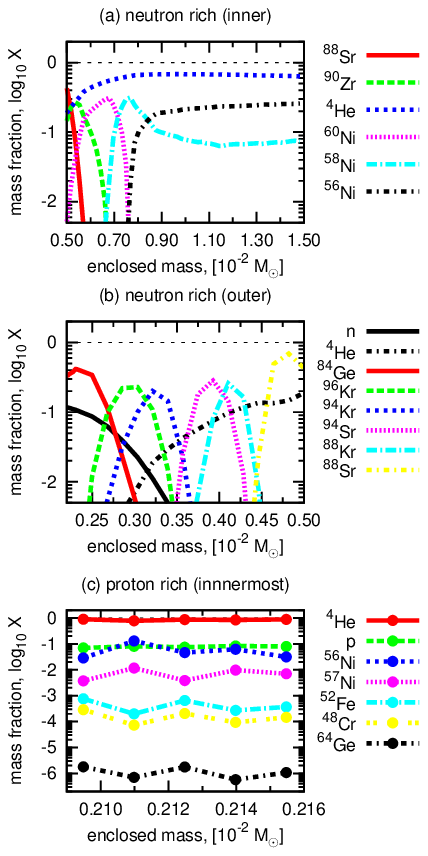}
		\caption{
Overview of dominating conditions, i.e. abundances of a few key nuclei
after charged-particle freeze-out, as a function of radial Lagrangian
mass, with variations from (a) typical explosive Si-burning products and an 
alpha-rich freeze-out in the outer zones to (b) an increasing remaining neutron
abundance after charged-particle freeze-out, permitting a weak $r$-process, down 
to (c) proton-rich neutrino-driven wind ejecta, permitting the onset of an 
$\nu$p-process.}
\label{fig-overview}
\end{center}
\end{figure}

The reheating leads to partial photodisintegration of heavy elements,
the production of alphas, and the buildup toward nuclei
with mass numbers around $A=90$--$110$ during the final expansion.
The neutron/seed ratio at this second charged-particle freezeout (boost)
is rather a measure for local rearrangements of matter.

The maximum neutron/seed ratio of slightly above 10 obtained over the complete range
of ejected mass zones, does not support conditions to produce the third $r$-process
peak, in fact only a small production of the second peak is expected.
The effect of the ``boost", i.e. second reheating and expansion in mass zones 020--040,
rearranges/reshapes the abundance distribution via photodisintegrations and captures,
but does not alter this conclusion. 

With this background we have a look at the final abundances of representative
mass zones, displayed in Figure \ref{fig-finab}. We see in the outer layers
remaining alpha-fractions beyond 60\%. These are regions, which experience
entropies of $S = 33$--$50 k_b$/baryon and $Y_e$'s close to 0.5 and see vanishing
neutron/seed ratios. Thus, we expect essentially the production of the 
Fe-group up to $A=50$--$70$.
This can be observed in the last two subfigures
of Figure~\ref{fig-finab} (zones 080 and 120). Mass zones 60 and 70, which experience the highest 
entropies and moderately decreased $Y_e$'s, can move matter up to and (slightly) beyond
$A=90$ (for mass zone 70 still with a large fraction of matter in the Fe-group), 
mostly due to a more neutron-rich (lower $Y_e$) charged-particle freeze-out and not due to
further neutron processing (see neutron/seed ratio in Figure~\ref{fig-Xalnseed}). 
One can also see a variation in the final 
carbon-fraction, underlining how effective the bridging of the alpha-to-carbon
bottle neck of reactions is in comparison to subsequent capture reactions
to heavier nuclei.
At smaller radii (mass zones 020 to 051 and $Y_e$'s as small as 0.33),
also the $A = 130$ peak starts to be populated. This is due to the
neutron/seed ratio of up to 10 attained after charged-particle freezeout. 
For the mass zones 20--45, abundances are shown after the first expansion
(dashed) and after reheating/boost (solid), which is characterized by some 
photo-disintegration of heavy nuclei, the appearance of $^4$He, and further 
processing of nuclei beyond the Fe-group (see also Figure~\ref{fig-boost}).
Summarizing the results of all mass zones with $Y_e < 0.5$, we note that
none of these zones produce matter beyond $A \sim 130$, nor show the 130
peak dominating abundances. 

Deeper mass zones, ejected via the NDW turns proton-rich
and they experience entropies as high as 
$85 k_B$ per baryon. These are conditions where we expect a $\nu p$-process
\citep{Froehlich:etal:2006b,Pruet:etal:2006,Wanajo:2006}.
This can be seen in the first three panels of Figure~\ref{fig-finab}, dashed lines show abundances
without the inclusion of the $\nu p$-process, solid lines show the final 
results after $\nu p$-processing.
While in terms of total mass fraction the production of nuclei beyond the Fe-group
is not too impressive, Figure~{\ref{fig-nup}} displays this more prominently, where the overproduction
ratio over solar is plotted. In the proton-rich environment
anti-neutrino captures on free protons produce neutrons and permit to overcome
the $rp$-process waiting point $^{64}$Ge via an $(n,p)$-reaction, winning against
a slower $\beta^+$-decay. In this way, nuclei up to $A = 80 - 90$ can be produced on
the proton-rich side of stability. The comparison of the black and red bullets 
indicates the strength of this process.

\begin{figure}[htbp]
	\begin{center}
		\includegraphics[width=\hsize]{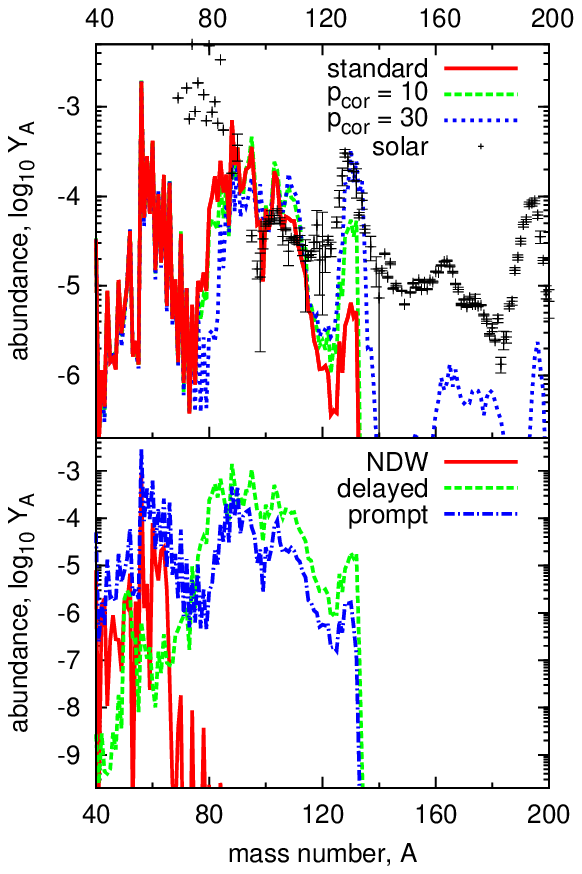}
		\caption{Integrated abundance distributions as a function of atomic
                mass number $A$ for all ejecta in comparison to solar r-abundances
                {\it (top)} (normalized to the A=100 region) and 
                separated by ejection process{\it (bottom)}.
                We also indicate the effect of uncertainties in the $Y_e$ determination,
                given in terms of percentage $P_{cor}$, discussed in section 5.}
\label{fig-integ}
	\end{center}
\end{figure}

\section{Survey and integrated composition}
After having discussed the individual composition, ejected from different 
positions in the exploding model, we want to give a final survey of the conditions 
attained in all mass zones of explosive Si-burning which are affected by
the explosion mechanism (here the deconfinement phase transition). The discussion
of nucleosynthesis in layers further out is only affected by the energy in the
shock wave, as has been discussed extensively in the literature
\citep{Thielemann:etal:1996}, and will not be repeated here. Finally, we also 
discuss the overall features of the integrated yields (of these inner mass zones, 
close to the explosion mechanism). 
The survey is displayed in Figure~{\ref{fig-overview}} which features abundances
after charged-particle freeze-out, and thus the setting for the final
nucleosynthesis features, if free neutrons or protons are remaining in sizable
fractions. 

\begin{figure}[htbp]
	\begin{center}
		\includegraphics[width=\hsize]{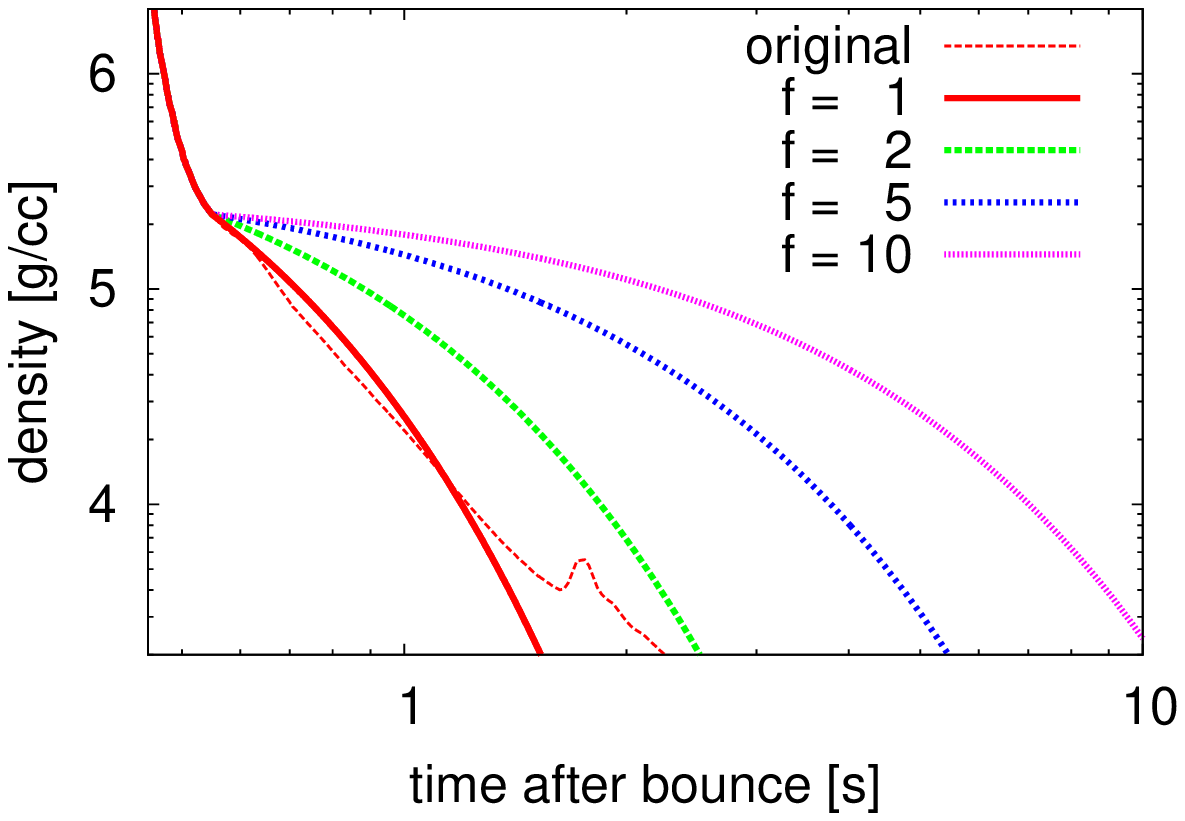}
		\caption{Density evolution as a function of time.
Shown is the standard case with the original explosion simulation according to EOS2 and
exponential density expansions fitted to this case ($f=1$) as well as slower expansions with
longer expansion timescales by a factor 2, 5, and 10.}
\label{fig-dens}
	\end{center}
\end{figure}

Essentially, all mass zones have total entropies in
excess of $S=30 k_B$ per baryon. In the outer mass zones with a $Y_e$ close to
0.5 this produces high alpha-fractions plus dominantly $^{56}$Ni. Moving
somewhat further in with slightly increasing entropies, noticeable amounts of
$^{64}$Ge are produced as well, with the $^{64}$Ge/$^{56}$Ni being a measure
of entropy. Decreasing $Y_e$, i.e., having more neutron-rich conditions, changes 
the initially pure alpha-rich charged-particle freezeout to an alpha-rich freeze-out
with sizable abundances up to $A=90$. A further decrease in $Y_e$ causes 
remaining free neutrons, which permit an additional sequence of neutron
captures. The decrease of $Y_e$ down to $0.33$, leads to a charged-particle 
freeze-out with vanishing alpha-fractions but a sufficient amount of free neutrons,
which permit later to produce nucleosynthesis ejecta in the mass A=130 peak.
The mass zones which are affected by a reheating boost are characterized by 
photo-disintegrations and a second charged-particle freeze-out with remaining
alpha-fractions, 
Finally the innermost proton-rich zones, with higher entropies of about 
$S=85 k_B$ per baryon and experiencing a
continuous neutrino-flux, show an alpha-rich and proton-rich freeze-out with
$Y_e$'s up to $0.55$. This causes a $\nu p$-process after charged-particle freeze-out
, permitted by neutron production via anti-neutrino captures on 
free protons. This process will produce  nuclei up to $A=90$, but on the proton-rich side
of stability, especially Sr, Y, and Zr isotopes.

After this survey of the conditions at charged-particle freeze-out, combined with
the final ejecta composition as a function of radial mass coordinate as discussed in
the previous section, we want to give an integrated presentation for these inner mass 
zones which experience conditions for the possible formation of nuclei beyond the Fe-group.
Figure~\ref{fig-integ} (top) shows the composition
for this range in mass numbers in comparison to solar $r$-abundances.
As we do not yet know the frequency of such events, i.e. which range of
the initial mass function of stellar masses leads to these types of explosions,
we show a scaling normalized to the $A \sim 100$ mass region (where the dominant
abundances are obtained).
The bottom part of Figure~\ref{fig-integ} shows the individual contributions 
to the overall abundances by the different mass zones as presented in Table \ref{tab-properties}
and Figure~\ref{fig-init}, and shown in different colors in Figure~\ref{fig-properties}
and \ref{fig-postprocess} as well as \ref{fig-entYe}, \ref{fig-Xalnseed}, and \ref{fig-finab}. 
The high end of the abundance distribution in the prompt (blue, dashed) as well as the 
delayed (boosted) ejecta (green, solid) are due to the most neutron-rich conditions with 
$Y_e$ close to $0.33$. 
The outermost ejected mass zones with $Y_e$ closer to $0.5$ contribute to the Fe-group
and matter up to $A = 80$ (lower range of mass numbers of blue dashed line).
The NDW (red line) with slightly proton-rich ejecta ( $Y_e = 0.55$) 
produces significant abundances only up to $A = 64$, if the same normalization for the
relevant mass zones is used (see, however, also Figure~\ref{fig-nup}).

\begin{figure}[htbp]
	\begin{center}
		\includegraphics[width=\hsize]{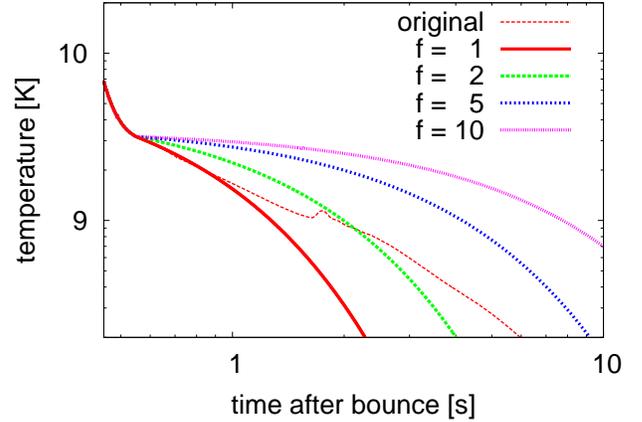}
		\caption{The corresponding temperature evolutions with adiabatic index 4/3.}
\label{fig-temp}
	\end{center}
\end{figure}

It is clearly visible that, first of all, objects like these supernovae, 
exploding
by a mechanism based on EOSs with a low-density quark--hadron
phase transition, do only experience a weak $r$-process in ejected mass zones 
which were neutronized during collapse.
There is no matter produced in the third $r$-process peak.
If normalizing the abundance curve at $A=100$, in order to avoid an overproduction 
of this mass region, also only a small contribution (less than 10\% is expected
to the second $r$-process peak, i.e., $A=130$). The major production affects the
atomic mass range from $A=80$ to $115$, curiously also reproducing a minimum 
at $A=97$--$99$. 
Thus, the type of events discussed here, contribute to the whole mass region
beyond the Fe-group up to $A=115$ in a significant way, accompanied by minor
contributions to the second $r$-process peak at $A = 130$.

\section{Model Uncertainties}

\subsection{Equation of State Uncertainties}
The major uncertainty in the present calculations is expected from the uncertain 
properties of the EOS, especially the quark--hadron phase transition which is at the
origin of the explosion mechanism discussed here. This was already reviewed extensively
in Section 2.1. The analysis done by \citet{Fischer:etal:2011} with variations in the
properties of the phase transition, led to different post-bounce times for the onset of the explosion
and different neutron star structures. Later transitions at higher densities cause more
massive progenitors with steeper density gradients at the surface.
The standard case of the present paper corresponds to EOS2 in Figure~17 of \citet{Fischer:etal:2011},
i.e., a late(r) explosion with steep(er) density gradient and thus a fast expansion. Variations
toward the properties of EOS1, with an earlier explosion, would cause slower expansions. Therefore, 
we performed test (nucleosynthesis) calculations with slower expansions for a specific mass zone
(here zone 51, see also Figures~\ref{fig-finab}) which provided among the best $r$-process conditions in our standard case.
Figs.{\ref{fig-dens}} and {\ref{fig-temp}} show variations with exponential density expansions
fitted to the standard case ($f=1$) and slower expansions with longer expansion times scales
by a factor of two, five, and 10 (choosing the temperature according to adiabatic expansions with
the same entropy).

\begin{figure}[htbp]
	\begin{center}
		\includegraphics[width=\hsize]{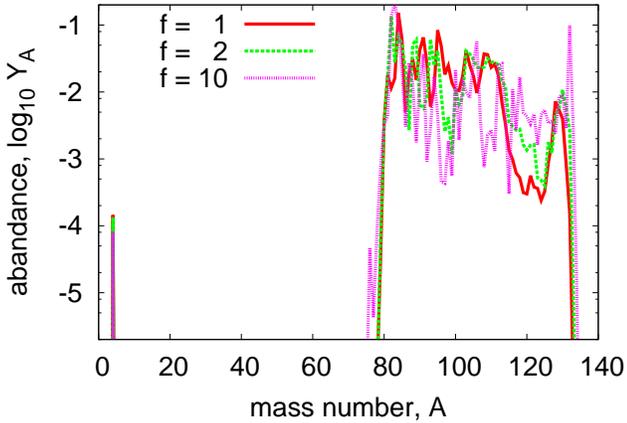}
		\caption{
		Final abundances for mass zone \# 51 with different expansion speeds.
	The slower expansions ($f=2, 10$) lead to smaller remaining alpha fractions (as expected), but
	also to a slightly stronger (weak) $r$-process, indicated by the abundances in the $A=130$ peak.}
\label{fig-fabund}
	\end{center}
\end{figure}

In typical NDW environments for given $S$ and $Y_e$, a fast(er) expansion leads to
a strong(er) alpha-rich freeze-out, thus more alpha particles and less seeds and a higher 
neutron/seed ratio, promoting the options for a stronger $r$-process. However, in the mass zones of 
interest here, the $Y_e$ is quite low and the remaining alpha-fraction close to negligible.
This feature was discussed extensively in Section 3.3
with respect to Figures~\ref{fig-Xalnseed} and \ref{fig-finab}.
We see the effect in Figure~\ref{fig-fabund}, that the fastest expansion ($f=1$) shows the highest
remaining alpha-fraction, but only of the order $10^{-4}$. This behavior lowers the initial seed
abundance after charged-particle freeze-out by the same amount, which is negligible. 
This causes essentially the same seed abundances and neutron/seed ratio for all calculations with 
different expansion speeds ($f=1$ -- $10$) and cancels this dominant effect in typical neutrino-driven wind
environments. In fact, the slowest expansion leads to the highest abundances at $A=130$.
Here, a second-order effect seems to take over, the different density
and temperature conditions for slower expansions 
lead to an apparently slightly more neutron-rich $r$-process path, encountering
smaller beta-decay half-lives which permit to proceed faster to heavier nuclei. This behavior,
as shown in Figure~\ref{fig-fabund}, indicates a slightly stronger (weak) $r$-process for the slower
expansions. However, the effect is not strong, only clearly visible for $f=10$, which is larger
than the uncertainty encountered in our EOS variations.}

\subsection{$Y_e$ Uncertainties}

One can argue, that there might exist some uncertainty in weak interactions
\citep[electron/positron captures and neutrino captures; see, e.g.][]
{LangankePinedo:2003, Janka:etal:2007, Cole:etal:2012}
which determine $Y_e$,
especially due to recent developments of including charged-current rates
that are consistent with the EOS
\citep[for details, see][]{MartinezPinedo:2012,Roberts:2012}.
These very recent preliminary investigations include medium effects for the (electron) neutrino
and antineutrino capture reactions, which lead to changed reaction $Q$-values
by adding nucleon interaction potential differences for neutrons and protons, enhancing as a 
result the difference between average anti-neutrino and neutrino energies. This causes a change in
$Y_e$. All our earlier discussions had the main emphasis that the influence of the neutrino
flux from the proto-neutron star would cause an increase in $Y_e$. The preliminary analysis
of such effects by \citet{MartinezPinedo:2012} and \citet{Roberts:2012}, which were not yet
included in the present calculations, shows that the $Y_e$ increase is weakened or even moderately
neutron-rich conditions ($Y_e \lesssim 0.5$) can result.

\begin{figure}[htbp]
	\begin{center}
		\includegraphics[width=\hsize]{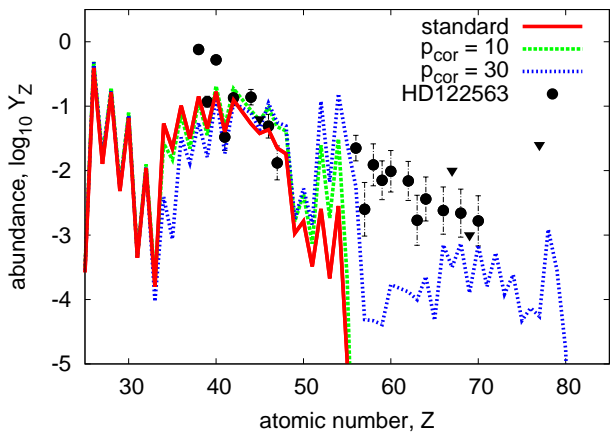}
		\caption{Integrated abundance distributions as a function of atomic
                charge number $Z$ in comparison to abundance features in "weak $r$-process"
                low metallicity stars as observed by \citet{Honda:etal:2006}.
                We also indicate the effect of uncertainties in the $Y_e$ determination,
                given in terms of percentage $P_{\rm{cor}}$}
\label{fig-honda}
	\end{center}
\end{figure}

For this reason, we also repeated the present nucleosynthesis calculations
with variations in the initial $Y_e$ for the mass zones experiencing
prompt and delayed explosions, according to the following recipe:
\[ Y_{e,\; \rm{cor}} = 0.5 + (Y_e - 0.5) \times \left(1 + \frac{p_{\rm{cor}}}{100}\right) \ ,\]
where the ${Y_{e,\; \rm{cor}}}$'s are the corrected ones
and $p_{\rm{cor}}$ denotes the percentage of uncertainty in deviations of $Y_e$ from
the symmetric value $0.5$, which enlarges these deviation from $0.5$.
This lowers the initially only
moderately neutron-rich $Y_e$'s in regions which already produced $A=130$ nuclei. The nucleosynthesis
results are presented in Figures~\ref{fig-integ} and \ref{fig-honda} and show the options of 
obtaining a full or weak $r$-process.
However, we expect that any uncertainties beyond $20$\% are unrealistic
for the explosion model we adopt in the current work.

As is obvious from the discussion above, that only a ``weak" $r$-process can be supported by 
the nucleosynthesis conditions found in the explosion mechanism discussed and presented
here, one might wonder whether such conditions support abundance features found in ``weak 
$r$-process" low-metallicity stars as observed by \citet{Honda:etal:2006}. For this reason,
we also show such a comparison in Figure~\ref{fig-honda}. What can be seen is that these 
observations also show sizable $r$-process features above the $A=130$ peak, although weaker
than in solar $r$-element abundances. If such abundance distributions are the result of a single 
nucleosynthesis pollution, also the ``weak" $r$-process found in the present paper can
only marginally explain such features. One could argue, however, that such observed abundance features
are a combination of at least two pollutions, one (low level) solar $r$-contribution plus
another ``weak" $r$-contribution extending only up to $A=130$.

\subsection{Multi-Dimensional Effects}

Nucleosynthesis uncertainties due to multi-dimensional effects, i.e., convective turnover,
have not been considered here, but we want to discuss their possible impact.
Note that convection and the presence of unstable fluid motion
are typically associated with compositional mixing, also of the ejected material.
Their development, e.g., after the shock passage across a low-density environment,
can leave peculiar features such as discussed in \citep{WanajoJanka:2012}
at the example of a $8.8 M_\odot$ star explosion.
In order to give an estimate for our nucleosynthesis results obtained
in the presence of possible convection,
we evaluated the Ledoux criterion \citep{WilsonMayle:1988}.
It is related to entropy per baryon and lepton-number gradients.
We find that regions where these gradients become largely negative,
i.e. regions where convection may develop, correspond to either fall-back
of material or the shock propagation in $Y_e\simeq0.5$ material.
Thus, matter with such uncertainties is either not ejected (fallback)
or mixing occurs in regions with very similar $Y_e$.
Hence, we expect the impact of possible convection
on the here discussed nucleosynthesis scenario to be of minor importance,
in particular the ejected material which attains low-$Y_e$ is not affected.
Independent of this expected unimportance of convection for nucleosynthesis results,
we should keep in mind that matter is strongly accelerated and thus ejected 
very fast. Even if instabilities exist, they should not lead to significant mixing
on such small timescales after the onset of the explosion and affect the nucleosynthesis.

Nevertheless, this nucleosynthesis site inhabits another possibility
of convection due to the explosion mechanism related to the quark--hadron phase transition.
It takes place on timescales on the order of several 100 ms,
depending on details of the quark--hadron EOS,
in the absence of any other, e.g., neutrino-driven explosion.
On these timescales, large-scale convection may develop
before the transition/explosion is initiated.
Exploring its impact on the entire explosion and nucleosynthesis scenario
will require multi-dimensional simulations, especially related to the point
in time when this explosion starts
(possibly affected by convection and related neutrino losses
and/or uncertainties in the EOS,
which both determine the $Y_e$ at the onset of the explosion).
This is beyond the scope of the present article and will require future investigations.

\section{Discussion and Summary}

Supernova nucleosynthesis is well understood for the outer ejected mass zones,
which can be well approximated by a shock wave with appropriate energy passing
through the layers of the progenitor 
\citep{WoosleyWeaver:1995,Thielemann:etal:1996,Woosley:etal:2002,
Nomoto:etal:2006,WoosleyHeger:2007,HegerWoosley:2010,Thielemann:etal:2011}.
What remains uncertain is the composition of the innermost ejecta,
directly linked to the explosion mechanism, i.e., the collapse and explosion phase.
In the present paper, we analyzed these mass zones of core-collapse supernova
explosions triggered by a quark--hadron phase transition during the early
post-bounce phase \citep{Sagert:etal:2009,Fischer:etal:2011}. 
A number of aspects are important for understanding these results.
The very innermost ejecta are strongly affected by the NDW.
Recent investigations noticed that this NDW turns matter
proton-rich, producing specific Fe-group isotopes and in the subsequent
$\nu p$-process nuclei with masses up to $A=80-90$
\citep{Liebendoerfer:etal:2003,Pruet:etal:2005,
Froehlich:etal:2006a,Froehlich:etal:2006b,Pruet:etal:2006,Wanajo:2006}. 
Even in the long-term evolution proton-rich
conditions prevail \citep{Fischer:etal:2010,Huedepohl:etal:2010}.
Thus, there seems to exist no chance to produce $r$-process matter in these
innermost regions, despite many interesting parameter studies for NDW
ejecta in terms of entropy $S$, electron fraction $Y_e$, and expansion
timescale $\tau$
\citep{Hoffman:etal:1997,MeyerBrown:1997,Freiburghaus:etal:1999,Farouqi:etal:2010}
or hydrodynamic studies, partially with parameter variations
\citep{Arcones:etal:2007,Kuroda:etal:2008,PanovJanka:2009,Roberts:etal:2010,ArconesMontes:2011}. 
Very recently, \citet{MartinezPinedo:2012} and \citet{Roberts:2012}, explored charged-current weak 
processes (i.e., electron-neutrino and anti-neutrino captures on neutrons and protons), consistent
with the EOSs in neutrino-driven supernova explosion. They find modifications of the
reaction $Q$-values due to medium effects, which increase spectral differences between $\nu_e$ and
$\bar\nu_e$. This results in slightly neutron-rich ejecta.
Nevertheless, since the entropy per baryon is still low a strong $r$-process
is unlikely to occur.

In order to obtain $r$-process conditions, a better chance to eject neutron-rich
matter is provided, when neutron-rich matter stems from the initial collapse and
compression, where electron captures made it neutron-rich, early in an explosion,
before neutrino interactions have the chance to turn it proton-rich in the NDW.
Core-collapse supernovae, exploding via the quark--hadron phase
transition (the focus of the present study), or electron-capture
supernovae \citep{Wanajo:etal:2011}, which explode without a long phase of
accretion onto the proto-neutron star, both lead to a rather prompt ejection of
prior compressed and neutronized matter.
However, the $Y_e$ obtained under such conditions
does not support a full $r$-process.
This kind of outcome also characterizes the conditions we find in the prompt
and quasi-prompt ejecta of the present study, which did not experience a strong
neutrino flux.
However, the $Y_e$-values attained are not smaller than $0.33$.
Whether uncertainties in the input or explosion physics can change this
down to values close to $0.23$, necessary for obtaining the third $r$-process peak,
remains to be shown in an upcoming article where we will include 
charged-current reaction rates that are consistent with the EOS
following \citet{MartinezPinedo:2012} and \citet{Roberts:2012}.

A related, but different, phenomenon has been discussed
in \citet[, and references therein]{Jaikumar:2007}.
In their scenario, the quark--hadron phase transition does
not happen shortly after core collapse and is not the cause of the supernova explosion.
Instead a regular neutrino-driven supernova explosion occurs first and leads to a
deleptonized neutron star with low-$Y_e$ and a steep density gradient at the surface.
Their scenario of a ``quark-novae", occurring via a quark--hadron phase transition in
an existing neutron star, is reported to produce $r$-process elements via the prompt 
ejection of extremely neutron-rich material from the neutron star surface.
In this sense, quark-novae differ from the scenario explored in the current study.
They result in a more compact object already in $\beta$-equilibrium, such that
the expansion timescale of the ejected material is much faster, compared to
the scenario discussed in the present paper.
However, in \citet{Jaikumar:2007} the authors apply a very simplified description
for the quark--hadron phase transition and ignore neutrino transport in general,
as well as weak processes in the hadronic part of their neutron star.
The latter may be essential for producing neutrinos which may change $Y_e$
during shock passage and hence lead to different nucleosynthesis results than
what has been discussed in \citet{Jaikumar:2007}.
It remains to be shown if the obtained favorable conditions for the $r$-process
will remain, applying more sophisticated input physics.

Given our results described above, we conclude that supernova explosions
triggered via the quark--hadron phase transition during the early post-bounce
phase can contribute to a weak $r$-process, consistent with observations
of low-metallicity stars \citep{Honda:etal:2006} and with LEPP abundances
\citep{Travaglio:etal:2004}.
Such conditions apparently do not occur in regular core-collapse supernovae.
However, strong $r$-process conditions, which also produce the third $r$-process 
peak and the actinides, have in simulations only materialized in 
neutron star mergers \citep{Freiburghaus:etal:1999b,Goriely:etal:2011},
fast rotating core-collapse supernovae with strong magnetic fields and
jet ejecta \citep{Cameron:2003,Nishimura:etal:2006,Fujimoto:etal:2008} or 
accretion disks around black holes \citep{Surman:etal:2008,WanajoJanka:2012}.
For the explosion model discussed here, investigation of uncertainties
in the hydrodynamics and the nuclear physics input showed that
a strong $r$-process to produce up to the third peak elements is rarely possible to obtain.

Neutron star mergers have been shown to be powerful sources of $r$-process matter,
in fact ejecting a factor of $100$ -- $1000$ more $r$-process material than required
on average from core-collapse supernovae, if those would have to explain solar
$r$-process abundances.
This would actually support the large scatter of Eu/Fe in comparison to,
e.g., O, Mg, Si, S, and Ca/Fe, where the latter are clearly produced in supernovae.
The only problem is that it might be hard to explain the early appearance
of $r$-process matter for metallicities at and below $\rm{[Fe/H]}=-3$.
Neutron star mergers will appear after the first supernovae
have already produced Fe and studies by \citet{Argast:etal:2004} showed
that one expects $r$-process products only for metallicities
$\rm{[Fe/H]} = -3 \sim -2$.
Some recent studies, which include the fact that our Galaxy is possibly the
result from smaller merging subsystems (with different star formation rates)
have been expected to show a way out of this dilemma.
If this cannot solved, we need another strong $r$-process source already
at low metallicities, and possibly jets from rotating core collapses with strong
magnetic fields could be the solution
\citep[see, e.g,][]{Nishimura:etal:2006,Fujimoto:etal:2007,Winteler:2012}. 

\section*{Acknowledgments}

We are grateful to the anonymous referee for the valuable comments
and suggestions to improve our manuscript.
N.N. acknowledges M. Hashimoto and S. Fujimoto for early development of
the nuclear reaction network which is extended and used in this work
and also thanks for A. Arcones to critical and useful comments.
The project was funded by the Swiss National Science Foundation
grant Nos. PP00P2-124879/1 and 200020-122287,
the Helmholtz Research School for Quark Matter Studies,
and the Helmholtz International Center (HIC) for FAIR.
N.N. is supported by the National Astronomical Observatory of Japan
under ``FY2012 Visiting Fellowship Program".
T.F is supported by the Swiss National Science Foundation under
project No. PBBSP2-133378 and G.M.P is partly supported by the
Sonderforschungsbereich 634, the ExtreMe Matter Institute EMMI and HIC for FAIR.
C.F. acknowledges support from the DOE Topical Collaboration
``Neutrinos and Nucleosynthesis in Hot and Dense Matter"
under contract DE-FG02-10ER41677.
M.L. and M.H. acknowledge support from the High Performance and High
Productivity Computing (HP2C) project.
M.H. is supported by the Swiss National Science Foundation (SNF) under project No. 200020-132816/1
and is also grateful for participating in the ENSAR/THEXO project.
T.R. is supported by the European Commission under the FP7 ENSAR/THEXO project.
The authors are additionally supported by CompStar, a research
networking program of the European Science Foundation and EuroGENESIS, a 
collaborative research program of the ESF. F.-K.T. is an Alexander von 
Humboldt awardee.

\bibliographystyle{apj}

\end{document}